\documentclass[nofootinbib,showpacs,preprintnumbers,amsmath,amssymb]{revtex4-1}

\def\bes{\begin{subequations}}
\def\ees{\end{subequations}}
\def\be{\begin{equation}}
\def\ee{\end{equation}}
\def\bea{\begin{eqnarray}}
\def\eea{\end{eqnarray}}
\def\ba{\begin{eqnarray}}
\def\ea{\end{eqnarray}}
\def\bear{\begin{array}}
\def\eear{\end{array}}
\def\p1sl{\displaystyle{\not}p_1}
\def\p2sl{\displaystyle{\not}p_2}

\def\enu{e^\pm \nu}
\def\munu{\mu^\pm \nu}
\def\taunu{\tau^\pm \nu}

\def\tauN{\tau^\pm N}
\def\Btolnu{B^+ \to \ell^+ \nu}

\def\Btotaunu{B^\pm \to \taunu}
\def\BtolN{B^\pm \to \ell^\pm N}

\def\BtotauN{B^\pm \to \tauN}

\def\BtoDtauN{B^\pm \to D \tauN}

\def\BtoDstauN{B^\pm \to D^* \tauN}

\newcommand{\bG}{{\overline{\Gamma}}}

\newcommand{\Dst}{D^*}
\newcommand{\blam}{{\overline \lambda}}
\usepackage{mathtools}
\usepackage{graphics}
\usepackage{graphicx}
\usepackage{dcolumn}
\usepackage{bm}
\usepackage{epsfig}
\usepackage{graphicx}
\usepackage{multirow}
\usepackage{dcolumn}
\usepackage{graphicx,epsfig}%
\usepackage{color}
\usepackage{amsmath}
\usepackage{slashed}

\begin{document}
\preprint{}

\title{Anomalies in (semi)-leptonic $B$ decays $B^{\pm} \to \tau^{\pm} \nu$,$B^{\pm} \to D \tau^{\pm} \nu$ and $B^{\pm} \to D^{*} \tau^{\pm} \nu$, \\
and possible resolution with sterile neutrino}

\author{Gorazd Cveti\v{c}}
\email{gorazd.cvetic@usm.cl }
\affiliation{Department of Physics, Universidad T\'ecnica Federico
  Santa Mar\'ia, Valpara\'iso, Chile}

\author{Francis Halzen}
\email{francis.halzen@icecube.wisc.edu}
\affiliation{Wisconsin IceCube Particle Astrophysics Center and Department of Physics, University of Wisconsin, Madison, WI, 53706}

\author{C.~S.~Kim}
\email{cskim@yonsei.ac.kr, Corresponding Author}
\affiliation{Department of Physics and IPAP, Yonsei University, Seoul
  120-749, Korea}

\author{Sechul Oh}
\email{scohph@yonsei.ac.kr}
\affiliation{University College, Yonsei University, Incheon 406-840, Korea}

\date{\today}

\begin{abstract}
\noindent The universality of the weak interactions can be tested in semileptonic $b \to c$ transitions, and in particular in the ratios $R(D^{(*)}) \equiv \Gamma(B \rightarrow D^{(*)} \tau \nu )/\Gamma(B \rightarrow D^{(*)} \ell \nu )$ (where $\ell = \mu$ or $e$). Due to the recent differences between the experimental measurements of these observables by BaBar, Belle and LHCb \textcolor{black}{ on the one hand and the Standard Model predicted values on the other hand, we study the predicted ratios $R(D^{(*)}) = \Gamma(B \rightarrow D^{(*)} \tau +{\rm ``missing"} )/\Gamma(B \rightarrow D^{(*)} \ell \nu )$ in scenarios with
an additional sterile heavy neutrino of mass $\sim 1$ GeV.
Further, we \textcolor{black}{evaluate} the newly defined ratio $R(0) \equiv \Gamma(B \to \tau +{\rm ``missing"} )/\Gamma(B \to \mu \nu )$ in such scenarios, in view of the future possibilities of measuring the quantity at Belle-II.}
\end{abstract}

\pacs{}

\maketitle


\section{Introduction}
\label{intr}

The lepton universality of weak gauge theory can be tested in exclusive semileptonic $B$ decays, possibly through the existence of new charged currents, as well as the quark flavor mixing structure
\cite{Cabibbo:1963yz} of the Standard Model (SM). To achieve these goals, it is usually necessary to calculate accurately the corresponding hadronic matrix elements. However, in the ratios like
\begin{equation}
R(D^{(*)}) \equiv \frac{\Gamma(B \rightarrow D^{(*)} \tau \nu )}{\Gamma(B
  \rightarrow D^{(*)} \ell \nu )} \, ,
\end{equation}
with $\ell =e~\text{or}~\mu$, most of the hadronic uncertainties cancel, making such ratios particularly relevant for testing the universality of weak interactions.
In Table \ref{tb2} we show the SM predictions which we will use. The shown  (hadronic) uncertainties originate, respectively, from lattice calculations \cite{Na:2015kha} and from estimated higher order correction in HQET to the ratio $A_0/A_1$ of form factors \cite{Fajfer:2012vx}.
Recently, new improved experimental results from Belle \cite{Huschle:2015rga} and LHCb \cite{Aaij:2015yra} Collaborations appeared, in addition to the older BaBar results \cite{Lees:2012xj}. As a result, the world average values reported by the HFAG group\cite{HFAG:leptonic,HFAG} are larger than the SM predicted values for $R(D)$ by 1.9 $\sigma$, and for $R(D^*)$ by 3.3 $\sigma$ (cf.~Table  \ref{tb2}).
%
\begin{table}[h]
\caption{Experimental results~\cite{Huschle:2015rga,Aaij:2015yra,Lees:2012xj} and the SM predictions~\cite{Na:2015kha}  of $R(D)$ and $R(D^*)$. The SM prediction with the scalar form factor (SFF) variations \cite{Kim:2016yth}
is also given.
The first and second experimental errors are statistical and systematic, respectively. }
\label{tb2}
\smallskip
\begin{tabular}{c c c} \hline \hline
 &  $R(D)$  &  $R(D^*)$  \\
\hline
BaBar &  ~~$0.440 \pm 0.058 \pm 0.042$~~  &  ~~$0.332 \pm 0.024 \pm 0.018$  \\
Belle &  ~~$0.375 \pm 0.064 \pm 0.026$~~  &  ~~$0.302 \pm 0.030 \pm 0.011$  \\
LHCb & $-$  &  ~~$0.336 \pm 0.027 \pm 0.030$  \\
Experimental average \cite{HFAG:leptonic,HFAG} &  ~~$0.397 \pm 0.040 \pm 0.028$~~  &  ~~$0.316 \pm 0.016 \pm 0.010$ \\
\hline
SM Prediction \cite{Na:2015kha} & $0.300 \pm 0.008$ & $0.252 \pm 0.003$ \\
SM Prediction (SFF Variation) \cite{Kim:2016yth} & $0.335$  & $-$ \\
\hline \hline
\end{tabular}
\end{table}

Many theoretical explanations have been proposed to explain these indications of possible lepton universality violation, among them the charged scalar exchanges
\cite{Fajfer:2012jt},
vector resonances \cite{Buttazzo:2016kid} or a $W^\prime$ boson
\cite{He:2012zp,Freytsis:2015qca,Bhattacharya:2016mcc},
leptoquarks (or, equivalently, R-parity violating supersymmetry)
\cite{Freytsis:2015qca,Tanaka:2012nw,Calibbi:2015kma,Bhattacharya:2016mcc}.
Effects of exchange of on-shell sterile neutrinos have also been evaluated,
cf.~ \cite{Abada:2013aba,Cvetic:2016fbv}.
Explanation of the anomalies within an effective field theory approach with  dimension-6 scalar, vector and tensor operators appeared in
\cite{Biancofiore:2013ki,Calibbi:2015kma}.
It was shown in
\cite{Fan:2015kna}
that the perturbative QCD (pQCD) combined with lattice results reduces the difference from the experimental values.
\textcolor{black}{ The electroweak effects in $B$-anomalies were investigated in Ref.~\cite{Paradisi}.}

In Ref. \cite{Kim:2016yth}, the authors checked how robust are the SM predictions for $R(D^{(*)})$ ratios. In contrast to the predictions of the vector form factors for $B\to D\ell \nu$ decays, which have been determined well in measurements of the  branching ratios and $q^2$-distributions in light lepton channels, the scalar form factor (SFF) is measurable only in decays with $\tau$ leptons. Even small deviations of the SFF from the lattice values can bring the SM  prediction closer to the current measured values of  $R(D)$, as shown in Table \ref{tb2}.

As closely related to the decays, $B \rightarrow D^{(*)} \tau \nu$,
the branching fraction (BF) of $B^+ \to \tau^+ \nu$ decay\footnote{Throughout this paper, \textcolor{black}{ formulas can be applied also to
charge-conjugate modes.}}
was measured by Belle and BaBar~\cite{BtotaunuBFs}, as shown in Table II~\cite{PDG2014,HFAG:leptonic,HFAG}
The SM prediction for the decay BF is given by \cite{CKMfitter}
\be \label{BtotaunuSM}
{\cal B}_{\rm SM} (B^+ \to \tau^+ \nu) = (0.848_{-0.055}^{+0.036}) \times 10^{-4} ~.
\ee
This implies that if the measurement is
improved in future $B$-factory experiments such as
Belle-II~\cite{BelleIIwww}, the comparison can clarify whether new physics
scenarios are needed.
%

\begin{table}[h]
\caption{Branching fractions of $B^+ \to e^+ \nu, ~\mu^+ \nu, ~\tau^+ \nu$ in units of $10^{-6}$~\cite{PDG2014,HFAG:leptonic,HFAG}.}
\label{tb1}
\smallskip
\begin{tabular}{c c c c}
\hline \hline
$B^+ \to$  &  $e^+ \nu$  &  $\mu^+ \nu$  &  $\tau^+ \nu$  \\
\hline
BABAR  &  ~~$< 1.9$~~  &  ~~$< 1.0$~~  &  ~~$179 \pm 48$ \\
Belle  &  ~~$< 0.98$~~  &  ~~$< 1.7$~~  &  ~~$91 \pm 19 \pm 11$ \\
Experimental average  &  $<0.98$  &  $< 1.0$  &  $106 \pm 19$  \\
\hline \hline
\end{tabular}
\end{table}
%
As a probe of new physics beyond the SM, the leptonic decays $\Btolnu$ are very interesting. This is so because these decay rates can be evaluated very precisely, and even at the tree-level new physics effects may appear, e.g., contributions of charged Higgs \cite{Hou} in two-Higgs doublet models~\cite{HHG}. The leptonic decay rates of $\Btolnu$ are in the SM  proportional to the square of the charged lepton mass, $m_\ell^2$. Thus, the decays of $B^\pm$ to $\enu$ and $\munu$ are strongly suppressed in comparison with the decays to $\taunu$.
Here we define new ratio $R(0)$ \cite{Hou} as
\begin{equation}
R(0) \equiv \frac{\Gamma(B \rightarrow \tau \nu )}{\Gamma(B
  \rightarrow  \mu \nu )} \, ,
\end{equation}
which is one of the most interesting to test the
universality of weak interactions, since all the hadronic
uncertainties cancel in the ratio,
\textcolor{black}{ and the ratio is a function of $M_{\tau}^2/M_B^2$ (and
  $M_{\mu}^2/M_B^2$).}

Heavy sterile neutral particles (a.k.a. ``heavy neutrinos'') have
suppressed mixing with SM neutrinos and appear in
various new physics scenarios, among them the original seesaw
\cite{seesaw} with very heavy neutrinos, seesaw with neutrinos with mass
$\sim 0.1$-$1$ TeV \cite{1TeVNu}, or with mass $\sim 1$ GeV \cite{1GeVNu}.
\textcolor{black}{For some studies of the production of very heavy neutrinos with mass $\sim 100$ GeV at the LHC we refer to \cite{DasLHC}.}
We will include in our considerations the reactions
$\BtotauN, \BtoDtauN, \BtoDstauN$, where $N$ is any heavy sterile neutrino of the Dirac or Majorana type,
and interpret the measured branching fractions in the new
physics scenario. For example, even if $N$ is invisible in the detector, we can
still distinguish $\BtolN$ signals from $\Btolnu$ for $\ell = e$ or
$\mu$, because these are two-body decays and therefore the
momentum of the charged lepton in the $B$ meson rest frame is fixed by the mass of $N$.  However, in the case of decays $\BtotauN$, the produced $\tau^{\pm}$ particle decays fast and hence there are more than one neutrinos
in the final state, and the decay
signature of $\BtotauN$ cannot be distinguished from the
ordinary $\Btotaunu$.  Therefore, the experimentally observed signal of $\Btotaunu$ may include contributions from $\BtotauN$, and this signal we will denote as $B^{\pm} \to \tau^{\pm} + {\rm ``missing"}$.

Massive neutrinos $N$ mix in general with the standard flavor neutrinos, $e.g.$ as in a seesaw type new physics scenario.
We denote as $U_{\ell N}$ the mixing coefficient for the heavy mass eigenstate $N$ with the standard flavor neutrino $\nu_{\ell}$ ($\ell = e, \mu, \tau$).\footnote{ Other notations for $U_{\ell N}$
  exist in the literature, among others $V_{\ell 4}$ in
  \cite{Atre}; $B_{\ell N}$ in \cite{PilZPC}.}
The standard sub-eV
neutrino $\nu_{\ell}$ ($\ell =e, \mu, \tau$) can then be represented as
\be
\nu_{\ell} = \sum_{k=1}^3 U_{\ell \nu_k} \nu_k + U_{\ell N} N \ ,
\label{mix}
\ee
where $\nu_k$ ($k=1,2,3$) are the light mass eigenstates.
The $3 \times 3$ matrix $U_{\ell \nu_k}$ is the usual PMNS matrix \cite{Pmns}. In the relations (\ref{mix}) we assume the existence of only one additional massive sterile neutrino $N$, however, it can be extended with any number of $N$.
Then the extended (unitary) PMNS matrix $U$ would be in this case a $4 \times 4$ matrix,
 implying the relations
\be
\sum_{k=1}^3 |U_{\ell \nu_k}|^2 = 1 -|U_{\ell N}|^2 \ .
\label{unit}
\ee
One of our scenarios will be with this unitarity assumption. This will modify the decay width, due to the existence of a massive neutrino $N$,
by the amount $\Gamma(B^+ \to \tau^+ N) - \Gamma(B^+ \to \tau^+ N)|_{M_N=0}$ and
$\Gamma(B^+ \to D^{(*)} \tau^+ N) - \Gamma(B^+ \to D^{(*)} \tau^+ N)|_{M_N=0}$, where the minus terms are
due to the unitarity of $U$.

In the other scenario,
\textcolor{black}{the $3 \times 3$ PMNS mixing matrix is unitary, and}
 $N$ will be regarded as a neutral fermion which does not mix with the SM flavor neutrinos $\nu_{\ell}$, but couples with charged leptons such as $\tau$ in the same form as in the first scenario, for example,
  \be
  \Delta {\cal L} \sim {\widetilde g} {\bar \tau} \gamma^{\mu} {\widetilde W}^{-}_{\mu} N + {\rm h.c.} \Rightarrow
  \delta {\cal L} = \left( - \frac{g}{\sqrt{2}} \right) U_{\tau N} {\bar \tau} W^{-}_{X \mu} N + {\rm h.c.} \ ,
  \label{dL}
\ee
via mediation of a new physics charged gauge boson $\widetilde W^{\pm}$,
\textcolor{black}{ and subscript $X$ denotes either $L$ or $R$ (left or right-handed projection).}
Here $\widetilde W^{\pm}$ has the light SM gauge boson $W^{\pm}$ component, and this may lead to couplings (\ref{dL}), where the suppression effects of such (or similar) scenarios are parameterized in the parameter $U_{\tau N}$.
These couplings have the same form as in the previous scenario, but now we have no condition of unitarity (\ref{unit}).
And such violation of the unitarity manifests
unknown new physics beyond the SM.

For our analysis, we want to keep in a most generic form both the scenarios which lead to Eqs.~(\ref{mix})-(\ref{unit}) and those which lead to Eq.~(\ref{dL}). Nonetheless, we wish to mention, as a representative example for the mechanism of Eq.~(\ref{dL}), the LR-models \cite{LR1,LR2,LangSan}
\textcolor{black}{ with the gauge group $SU(2)_L \times SU(2)_R \times U(1)_{B-L}$.}
Such models have in the scalar sector a $2 \times 2$ LR-doublet $\phi$ and triplets $\Delta_L$, $\Delta_R$. The vacuum expectation value (VEV) $u_R$ of $\Delta_R$ is much larger than the other VEVs, $|u_R| \gg |v|, |w| \gg |u_L|$ (where $v, w \sim 10^2$ GeV are VEVs in $\phi$), leading to a hierarchical mixing of the charged flavor bosons ${\widetilde W}_L^{\pm}$ and ${\widetilde W}_R^{\pm}$. The flavor boson  ${\widetilde W}_R^{\pm}$ has, as a result, a small component ($\sim |v^2/u_R^2|$) of the SM mass eigenstate boson $W^{\pm}$.
\textcolor{black}{ More specifically,
\be
{\widetilde W}_R^{\pm} = - e^{\mp i \lambda} \sin \xi \; W^{\pm} + \cos \xi \; W_2^{\pm},
\label{WR}
\ee
where $W_2^{\pm}$ is the heavy mass eigenstate with mass $M_2^2 \approx g^2 |u_R|^2/4 \gg M_W^2$, $\lambda$ is a real phase, and the mixing angle $\xi$ is
\be
\xi \approx \frac{2 |v w|}{|v|^2+|w|^2} \left( \frac{M_W}{M_2} \right)^2,
\label{xi}
\ee
where $(|v|^2+|w|^2) = 1/(\sqrt{2} G_F)$. In such models we can have a heavy neutrino $N_R$ which forms with $\tau_R$ an $SU(2)_R$-doublet $(N_R,\tau_R)$ and $N_R$ is a flavor and mass eigenstate, i.e., it does not mix with other flavor neutrinos such as $(\nu_{\ell,L})^c$ ($\ell=e,\mu, \tau$) and $\nu_{\ell', R}$ ($\ell'=e,\mu$). Then the gauge boson mixing (\ref{WR}) gives us in the terms which couple ${\widetilde W}_R^{\pm}$ with leptons the following contributions (we take $g_R=g_L=g$):
\bes
\label{LRcontr}
\bea
- \frac{g}{\sqrt{2}} {\bar \tau} \gamma^{\mu} \left( \frac{1 + \gamma_5}{2} \right) N {\widetilde W}^{-}_{R \mu} + {\rm h.c.} & = &
  - \frac{g}{\sqrt{2}} {\bar \tau} \gamma^{\mu} \left( \frac{1 + \gamma_5}{2} \right) N \left( - e^{i \lambda} \sin \xi  \; W^{-}_{\mu} + \cos \xi \; W^{-}_{2 \mu} \right) + {\rm h.c.}
\label{LRcontra}
\\
& = & - \frac{g}{\sqrt{2}} U_{\tau N} {\bar \tau} \gamma^{\mu} \left( \frac{1 + \gamma_5}{2} \right) N  W^{-}_{\mu} + {\rm h.c.} + \ldots ,
\label{Rcontrb}
\eea
\ees
where the ellipsis stands for the couplings with the heavy $W_2^{\pm}$ boson. Here we see that in such a case the heavy-light mixing parameter is $|U_{\tau N}|^2 = \sin^2 \xi \approx \xi^2$, which, according to Eq.~(\ref{xi}) is then
\be
|U_{\tau N}|^2 \approx \left( \frac{M_W}{M_2} \right)^4.
\label{U2}
\ee
According to the analysis of general LR-models of Ref.~\cite{LangSan}, there is a lot of freedom in $V_R$, the right-handed quark mixing matrix, resulting in a relatively generous bound $M_2 > 300$ GeV coming predominantly from the $K_L$-$K_S$ mass difference $\Delta m_K$ and by $B_d {\bar B}_d$ mixing (and assuming $g_R=g_L$). This implies the upper bound $|U_{\tau N}|^2 < 5 \times 10^{-3}$ in such LR-scenarios.\footnote{If the LR-models are restricted to the minimal (symmetric) versions, where $V_R$ is closely related with $V_L$($ = V_{\rm CKM}$), the resulting bound is more restrictive, $M_2 > 2.3$ TeV \cite{NemSen}. The signals of $W_R$ were searched also at LHC (CMS Collaboration) \cite{CMSPAS}, in specific minimal LR-model scenarios where $W_R^{\pm}$ would decay in the $\ell=e, \mu$ channels to $\ell N_{\ell, R}$; mass exclusion regions $M_2 > 3.3$ TeV were found  \cite{CMSPAS} in such scenarios, but only if $M_{N_{\ell}} > 200$ GeV.}}

\textcolor{black}{ We further point out that the right-handedness of the coupling $\tau$-$W$-$N$, Eq.~(\ref{Rcontrb}), does not affect the formulas for the decay widths $\Gamma(B \to (D^{(*)}) \tau \nu)$ that we use in the present paper: we checked that these formulas turn out to be the same as in the case of the left-handedness of the $\tau$-$W$-$N$ coupling; we recall that the couplings of quarks to $W$ are, of course, always left-handed.}

In our numerical analysis, we will derive the results  for two scenarios: either the ($4 \times 4$) matrix $U$ is  unitary, or $U$ without unitarity assumption: $i.e.$ analyses with and without unitarity assumption.
Our formulas, to be derived in the following Sections, will be applicable also to cases with more than one additional massive neutral fermion $N$ where the second fermion has a mass $M_{N'} \gtrsim 10$ GeV, such particles being too heavy to be produced on-shell in the considered decays.

\textcolor{black}{ At present, the upper bounds for the mixing parameters $|U_{\tau N}|^2$ are available from the measurements and analyses of the CHARM \cite{CHARM} and DELPHI \cite{DELPHI} Collaborations. These were dedicated direct measurements and analyses for decays producing heavy neutrinos $N$.\footnote{\label{CHDEL} CHARM limits \cite{CHARM} were obtained for $M_N < 300$ MeV, based on the absence of signals $N \to \nu_{\tau} Z^0 \to \nu_{\tau} e^+ e^-$ in a neutrino beam dump experiment, where $N$ is produced from decays of $D_s$ mesons. DELPHI limits \cite{DELPHI} were obtained for $M_N > 250$ MeV, based on the absence of signals $e^+ e^- \to Z^0 \to \nu_{\ell} N$ for long-lived and short-lived $N$, and they apply to all three parameters $|U_{\ell N}|^2$ ($\ell=e, \mu, \tau$).}
  On the other hand, there are indirect indications that the heavy-light mixing parameters $|U_{\tau N}|^2$ have more restrictive upper bounds, coming from the $\tau$ lepton decays where the analyses were made under the assumption of the SM scenario of (practically) massless neutrinos and unitary $3\times 3$ PMNS matrix $U_{\rm PMNS}$, and there the lepton universality of the electroweak coupling $g$ was shown to a large precision \cite{HFAG} (Sec.~9.2 there). However, in these latter analyses, unlike in Refs.~\cite{CHARM,DELPHI}, it was assumed that heavy neutrinos do not exist. In view of the lack of any new updated dedicated measurements and analyses of the $\tau$ decays with heavy neutrinos, we will}
\textcolor{black}{usually present here the upper bounds on the mixing parameters $|U_{\tau N}|^2$ as those from Refs.~\cite{CHARM,DELPHI}.}
\textcolor{black}{Nonetheless, in the next Section we will present an analysis of the lepton universality results \cite{HFAG} in the scenario of one additional heavy neutrino $N$, and in the subsequent analyses in this work we will keep in mind the restrictions on the heavy-light mixing from such an analysis.}

In this paper, we will consider the recent experimental anomalies, $R(D)$ and $R(D^*)$,
with the theoretical assumption of one not very heavy neutrino $N$: $M_N \sim 1$ GeV.
We will also predict the newly defined $R(0)$, which can be measured at Belle-II, as a function of the unknown parameters, $M_N$ and $U_{\tau N}$.

\section{Restrictions on $|U_{\tau N}|^2$ from lepton universality tests in $\tau$ decays}
\label{sec:univ}

The Heavy Flavor Averaging Group (HFAG)  \cite{HFAG} obtained restrictions coming from the lepton universality tests of SM. They analyzed, among other things, the measured widths $\Gamma(\tau \to e + {\rm missing})$ and $\Gamma(\mu \to e + {\rm missing})$, where ``missing'' stands for $\nu_{\tau} {\bar \nu}_e (\gamma)$ and $\nu_{\mu} {\bar \nu}_e (\gamma)$. They thus obtained
\be
\left( \frac{g_{\tau}}{g_{\mu}} \right) =  1.0010 \pm 0.0015 \ ,
\label{HFAGUniv0}
\ee
where the above notation stands for
\be
\left( \frac{g_{\tau}}{g_{\mu}} \right)^2 = \left( \frac{M_{\mu}}{M_{\tau}} \right)^5 \frac{\Gamma(\tau^- \to \nu_{\tau} e^- {\bar \nu}_e (\gamma))}{\Gamma(\mu^- \to \nu_{\mu} e^- {\bar \nu}_e (\gamma))} \times \left[ \frac{R_{\gamma}(\mu)}{R_{\gamma}(\tau} \frac{R_W(\mu)}{R_W(\tau)} \frac{f(M_e^2/M_{\mu}^2)}{f(M_e^2/M_{\tau}^2)} \right] = 1 + (2 \pm 3) \times 10^{-3}.
  \label{HFAGUniv}
\ee
The above ratio in their analysis includes the QED ['$(\gamma)$'] \cite{MS} and other corrections [$\sim (M_{\tau}/M_W)^2$, $\sim (M_e/M_{\mu})^2$]. However, the net contribution of all these effects to the above quantity, represented as the correction factor in the brackets in Eq.~(\ref{HFAGUniv}), turns out to be $\sim 10^{-4}$ and will thus be ignored in the analysis here.

If, on the other hand, we have an additional heavy neutrino $N$ which couples to $\tau$ (but not to $e$ or $\mu$), either in a scenario in which the $4 \times 4$ matrix $U$ is nonunitary [e.g., scenarios as in Eq.~(\ref{dL})], or in a scenario where the $4 \times 4$ mixing matrix $U$ is unitary [cf.~Eq.~(\ref{unit})], the above analysis changes significantly.

Namely, in the scenario where the formal $4 \times 4$ matrix $U$ is nonunitary (and the $3 \times 3$ PMNS mixing matrix is unitary), the accounting for the nonzero mass $M_N$ changes the above analysis in the following way. The quantity $\Gamma(\tau^- \to \nu_{\tau} e^- {\bar \nu}_e)$ gets replaced by
\bes
\label{repl}
\bea
\Gamma(\tau^- \to \nu_{\tau} e^- {\bar \nu}_e) &\mapsto&  \Gamma(\tau^- \to \nu_{\tau} e^- {\bar \nu}_e)  + \Gamma(\tau \to N e^- {\bar \nu}_e)
\label{repla}
\\
& = &  \Gamma(\tau^- \to \nu_{\tau} e^- {\bar \nu}_e)  + |U_{\tau N}|^2 \bG(\tau \to N e^- {\bar \nu}_e),
\label{replb}
\eea
\ees
where $\bG(\tau \to N \ell^- {\bar \nu}_{\ell})$ stands for the decay width to a (massive) neutrino $N$ and the coupling parameter $|U_{\tau N}|^2=1$
\bea
\bG(\tau \to N \ell^- {\bar \nu}_{\ell}) &=& \frac{G_F^2}{96 \pi^3} M_{\tau}^5
\int_{z_{\ell}}^{(1 - \sqrt{z_N})^2} dz \; \lambda^{1/2}(1, z_N, z)  (z - z_{\ell})
\nonumber\\
&&\times \left\{ (1 + z_N -z) \left( 1 - \frac{z_{\ell}}{z} \right)^2 +
\left[ (1 - z_N)^2 \frac{1}{z} - z \right] \left[ 1 + \frac{z_{\ell}}{z} - 2 \left( \frac{z_{\ell}}{z} \right)^2 \right] \right\}.
\label{bG}
\eea
Here, $z_N = (M_N/M_{\tau})^2$, $z_{\ell} = (M_{\ell}/M_{\tau})^2$ ($\ell=e$ or $\ell = \mu$), and $z = p_W^2/M_{\tau}^2$ where $p_{W}^2$ is square of the invariant mass of $W^{*-} = (\ell {\bar \nu}_{\ell})$. The first term on the right-hand side of Eq.~(\ref{replb}) denotes the usual SM contribution
\be
\Gamma(\tau^- \to \nu_{\tau} e^- {\bar \nu}_e) = \sum_{k=1}^3 |U_{\tau \nu_k}|^2 \bG(\tau \to \nu_k e^- {\bar \nu}_e) = \bG(\tau \to \nu e^- {\bar \nu}_e)|_{M_{\nu}=0},
\label{SMGamma}
\ee
where the above identities hold because of the (practical) masslessness of the SM neutrinos $\nu_k$ ($k=1,2,3$) and the unitarity of the $3 \times 3$ PMNS matrix.

The following ratio is crucial in the present analysis:
\be
G_{\ell}(M_N) = \frac{\Gamma(\tau^- \to N \ell^- {\bar \nu}_{\ell})}{\Gamma(\tau^- \to \nu_{\tau} {\ell}^- {\bar \nu}_{\ell})} = |U_{\tau N}|^2 {\overline G}_{\ell}(M_N),
\label{GMN}
\ee
where $\ell=e$ or $\ell=\mu$, and
\be
{\overline G}_{\ell}(M_N) = \frac{\bG(\tau^- \to N \ell^- {\bar \nu}_{\ell})}{\bG(\tau^- \to \nu \ell^- {\bar \nu}_{\ell})|_{M_{\nu}=0}}
\label{bGMN}
\ee
is the corresponding canonical ratio.

\begin{figure}[htb]
\centering\includegraphics[width=100mm]{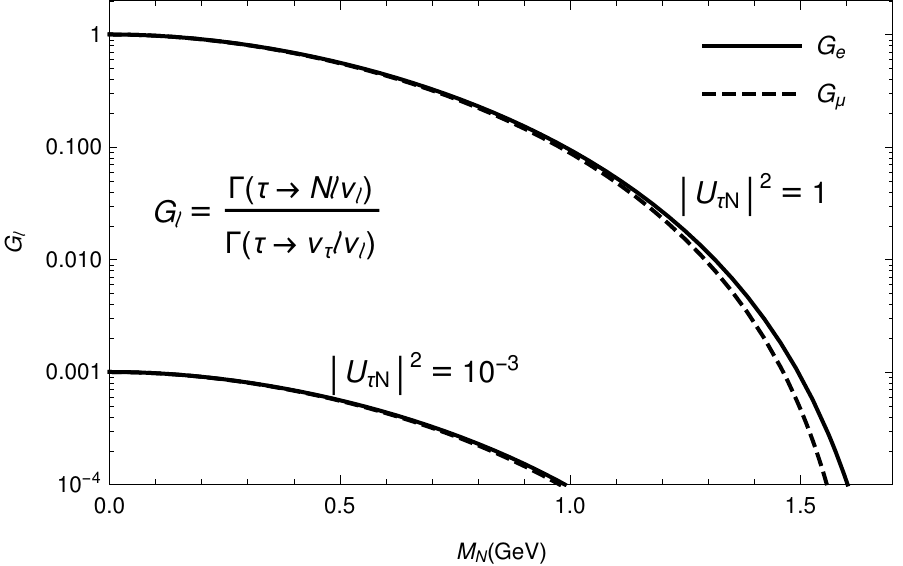}
\vspace{-0.2cm}
\caption{The ratios $G_{e}$ and $G_{\mu}$, Eq.~(\ref{GMN}), as a function of $M_N$, for two choices of the heavy-light mixing: $|U_{\tau N}|^2=1$ (then $G_{\ell} = {\overline G}_{\ell}$), and $|U_{\tau N}|^2=10^{-3}$.}
\label{figGMN}
\end{figure}

The values (\ref{HFAGUniv}), together with Eq.~(\ref{replb}) and the notations (\ref{GMN})-(\ref{bGMN}), then lead in the mentioned scenario to the following predictions for $|U_{\tau N}|^2$:
\be
G_e(M_N) = (2 \pm 3) \times 10^{-3} \Rightarrow
|U_{\tau N}|^2 = \frac{(2 \pm 3) \times 10^{-3}}{{\overline G}_{e}(M_N)}.
\label{UtauN2NU}
\ee
We recall that $N$ here is massive, on-shell, couples only to $\tau$ (and not to $\mu$ or $e$), and the $3 \times 3$ PMNS matrix is as in SM, i.e., unitary. The ratio function $G_{\ell}(M_N)$ is presented in Fig.~\ref{figGMN} for the cases $|U_{\tau N}|^2=1$ and $|U_{\tau N}|^2=10^{-3}$.

If, on the other hand, the $4 \times 4$ heavy-light mixing matrix $U$ is unitary, i.e., Eq.~(\ref{unit}), an analogous analysis leads to the conclusion that $|U_{\tau N}|^2$ has a very restrictive upper bound. Namely, in such a case the ratio on the left-hand side of Eq.~(\ref{HFAGUniv}) must be below the value of unity, and the most generous upper bound for $|U_{\tau N}|^2$ is then obtained if the right-hand side of Eq.~(\ref{HFAGUniv}) is $1 - 1 \times 10^{-3}$
\be
|U_{\tau N}|^2 < \frac{ (-1) \times 10^{-3}}{({\overline G}_e(M_N)-1)} = \frac{ (+1) \times 10^{-3}}{|{\overline G}_e(M_N)-1|}.
\label{UtauN2U}
\ee

\begin{figure}[h] 
\begin{minipage}[b]{.49\linewidth}
\centering\includegraphics[width=90mm]{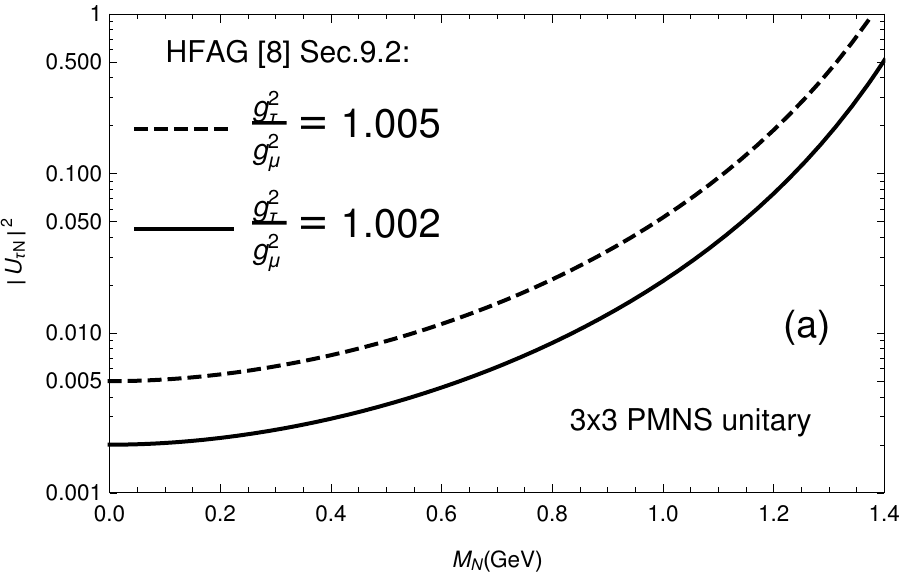}
\end{minipage}
\begin{minipage}[b]{.49\linewidth}
\centering\includegraphics[width=90mm]{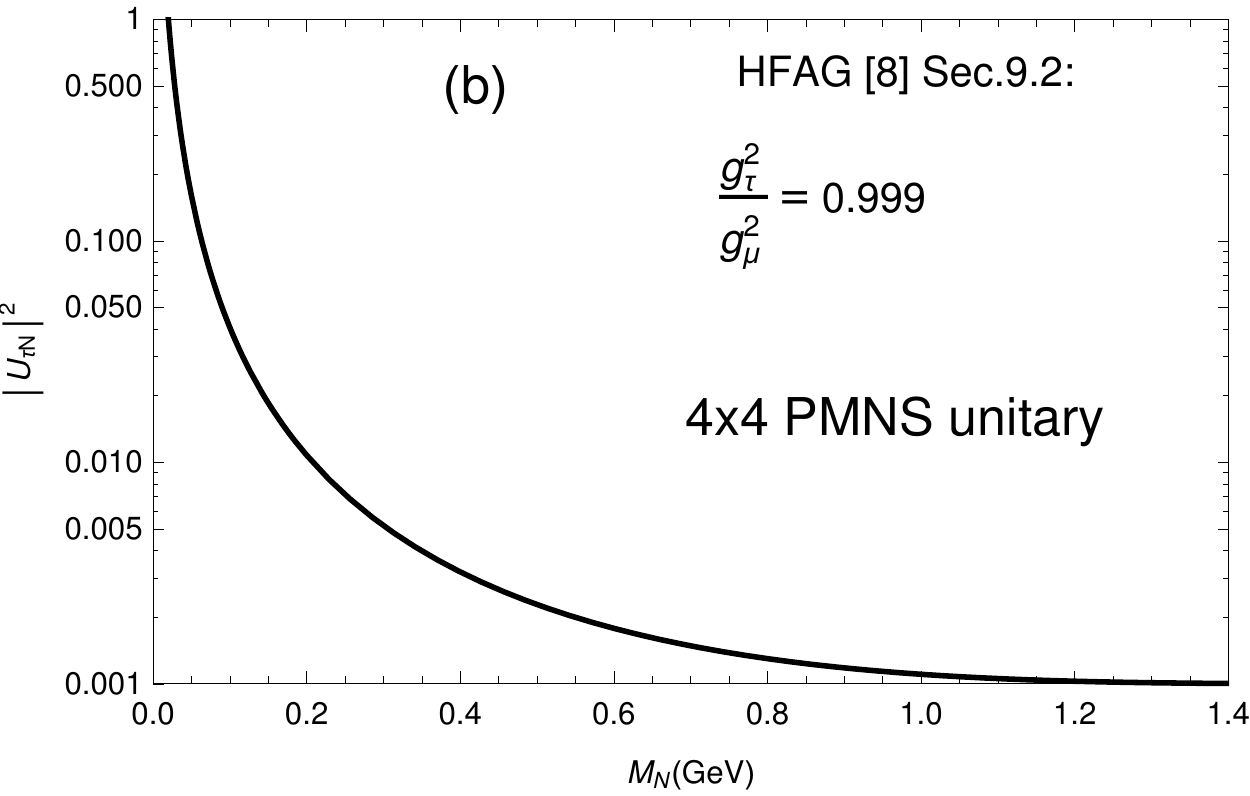}
\end{minipage}
\vspace{-0.2cm}
\caption{The upper bounds as obtained by lepton universality tests, Eq.~(\ref{HFAGUniv}): (a) in the case when the $3 \times 3$ PMNS mixing matrix is unitary (and the $4 \times 4$ matrix $U$ is nonunitary) - the solid line is the upper bound if the central value $(1+ 2 \times 10^{-3})$ is taken in the values Eq.~(\ref{HFAGUniv}), the dashed line when the largest value $(1+ 5 \times 10^{-3})$ is taken there; (b) in the case when the $4 \times 4$ mixing matix $U$ is unitary, and in that case the lowest value $(1 - 1 \times 10^{-3})$ is taken in Eq.~(\ref{HFAGUniv}).}
\label{figUtauN2Univ}
\end{figure}

The restrictions (\ref{UtauN2NU}) and  (\ref{UtauN2U}) are presented in Figs.~\ref{figUtauN2Univ}.

From Fig.~\ref{figUtauN2Univ} (a) we can see that in the scenario of the $3 \times 3$ unitary PMNS matrix (the $4 \times 4$ matrix $U$ is then nonunitary), the bounds (\ref{HFAGUniv}) imply for the heavy-light mixing parameter $|U_{\tau N}|^2$ stringent upper bounds $|U_{\tau N}|^2 \lesssim 0.5 \times 10^{-2}$ if the mass of $N$ is low ($M_N < 0.4$ GeV), and for higher masses the bounds are not stringent: $|U_{\tau N}|^2 \lesssim 10^{-1}$ for $M_N > 1$ GeV.

On the other hand, if the $4 \times 4$ matrix $U$ is considered unitary, Fig.~\ref{figUtauN2Univ} (b) implies that the upper bounds for $|U_{\tau N}|^2$ are quite stringent ($\lesssim 10^{-3}$) for higher masses $M_N> 0.6$ GeV, while for light masses less stringent upper bounds apply. The results of Fig.~\ref{figUtauN2Univ} (b) are consistent with the results of the analysis \cite{Becir} of the experimental constraints for the well-measured decay ratios ${\cal B}(\tau \to \ell \nu \nu')$ where $\ell=e, \mu$, and $\nu$ and $\nu'$ are light neutrinos or the heavy neutrino $N$, in the scenario which corresponds here to the unitary $4 \times 4$ $U$ matrix. The authors of Ref.~\cite{Becir} obtain the upper bound $|U_{\tau N}|^2 < 0.01$ for $0.3 \ {\rm GeV} < M_N < 1 \ {\rm GeV}$.

We will keep in mind these restrictions which come indirectly from the lepton universality tests Eq.~(\ref{HFAGUniv}), i.e., from the decays $\tau \to e + {\rm missing}$ and $\mu \to e + {\rm missing}$. Nonetheless, in the graphs we will account for the bounds on $|U_{\tau N}|^2$ coming from the dedicated direct measurements of the CHARM \cite{CHARM} and DELPHI \cite{DELPHI} Collaborations mentioned in the Introduction.


\section{Theoretical details of $R(D)$, $R(D^{*})$ and $R(0)$ with Light Sterile Neutrino}
\label{BtoDlN}

\subsection{$R(D)$ and decays of $B \to D \ell N$}
\label{sub:BtoDlN}

In the SM the amplitude for hadronic transition $B \to D$ is given in terms of vector and scalar form factors,
$F_1(q^2)$ and $F_0(q^2)$, defined as
\be \label{FFBD}
\langle D(p_D) | {\overline c} \gamma^{\mu} b | B^{-}(p_B) \rangle
 = \left[ (2 p_D + q)^{\mu} - \frac{(M_B^2-M_D^2)}{q^2} q^{\mu} \right] F_1(q^2)
  + \frac{(M_B^2-M_D^2)}{q^2} q^{\mu} F_0(q^2) \ ,
\ee
where $q=p_B-p_D$ is the momentum of the virtual $W^-$.
For the process $B \to D \ell N$, the decay width is given in Ref.~\cite{Cvetic:2016fbv} as
\bea \label{GBDlN}
&& \Gamma(B \to D \ell N) = |U_{\ell N}|^2 \; {\overline \Gamma}(B \to D \ell N) \ , \\
 {\overline \Gamma}(B \to D \ell N) &=& \frac{1}{384 \pi^3} G_F^2 |V_{c b}|^2 \frac{1}{M_B}
 \int_{(M_N +M_\ell)^2}^{(M_B -M_D)^2}  d q^2 \;
 \frac{1}{(q^2)^2}
 \lambda^{1/2} \left(1, \frac{q^2}{M_B^2}, \frac{M_D^2}{M_B^2} \right)
 \lambda^{1/2} \left(1, \frac{M_\ell^2}{q^2}, \frac{M_N^2}{q^2} \right)  \nonumber\\
&& \times
 {\bigg \{} F_1(q^2)^2 \left[ 2 (q^2)^2 -q^2 M_N^2 +M_\ell^2 (2 M_N^2- q^2) -M_N^4 -M_\ell^4 \right]
  \left[ (q^2 - M_D^2)^2 - 2 M_B^2 (q^2 + M_D^2) + M_B^4 \right]  \nonumber\\
&& + F_0(q^2)^2 3 (M_B^2 -M_D^2)^2 \left[ q^2 M_N^2 +M_\ell^2 (2 M_N^2 + q^2) -M_N^4 -M_\ell^4 \right]
     {\bigg \}} \ ,
 \label{bGBDlN}
\eea
where the kinematically allowed values of $q^2$ are $(M_N +M_\ell)^2 \leq q^2 \leq (M_B -M_D)^2$.
Notice that if $N$ is replaced by $\nu_\ell$ (i.e., $M_N \approx 0$), ${\overline \Gamma}(B \to D \ell N)$
in Eq.~(\ref{bGBDlN}) becomes the SM decay width of $B \to D \ell \nu_\ell$,
i.e., $\Gamma_{\rm SM} (B \to D \ell \nu_\ell)$.

The form factor $F_1(q^2)$ is well known~\cite{CLN}.\footnote{
  \textcolor{black}{For early attempts to account for the flavor symmetry breaking in form factors of heavy pseudoscalars, cf.~Ref.~\cite{Khlopov}.}}
It can be expressed in terms of the variable $w$
\bes
\label{wz}
\bea
w & = & \frac{(M_B^2 + M_D^2 - q^2)}{2 M_B M_D} \ ,
\label{w}
\\
z(w) & = & \frac{\sqrt{w+1} - \sqrt{2}}{\sqrt{w+1} + \sqrt{2}} \ ,
\label{z}
\eea
\ees
in the following approximate form~\cite{CLN}:
\be
F_1(q^2) = F_1(w=1) \left( 1 - 8 \rho^2 z(w) + (51 \rho^2 - 10) z(w)^2 - (252 \rho^2 - 84) z(w)^3 \right) \ ,
\label{CLNF1}
\ee
where the free parameters $\rho^2$ and $F_1(w=1)$ have been recently determined with high precision by the Belle Collaboration, Ref.~\cite{Belle1}
\bes
\label{rho2F1max}
\bea
\rho^2 &= & 1.09 \pm 0.05 \ ,
\label{rho2}
\\
|V_{cb}| F_1(w=1) &=& (48.14 \pm 1.56) \times 10^{-3} \ .
\label{F1max}
\eea
\ees
The value (\ref{F1max}) was deduced from their value of $\eta_{EW} {\cal G}(1) |V_{cb}| = \eta_{EW} F_1(w=1) \sqrt{4 r}/(1+r) = (42.29 \pm 1.37) \times 10^{-3}$, where $r=M_D/M_B$ and $\eta_{EW} =1.0066 \approx 1$~\cite{Sirlin}. In our numerical evaluations, we will use the central values $\rho^2 =  1.09$ and $|V_{cb}| F_1(w=1)  = 48.14 \times 10^{-3}$. A recent study~\cite{Kim:2016yth} has shown that the mostly unknown scalar form factor $F_0 (q^2)$, being expressed as
\be \label{F0_new}
F_0(q^2) = (1 + \alpha q^2 + \beta q^4) F_1 (q^2) ~,
\ee
can enhance the value of $R(D)$ \textcolor{black}{up to $0.335$ within} the SM, as shown in Table I.
We will use this scaling relation for $F_0 (q^2)$ with $\alpha = +0.16 ~ {\rm GeV^{-2}}$ and $\beta = -0.003 ~{\rm GeV^{-2}}$ as used in Ref.~\cite{Kim:2016yth}.

\textcolor{black}{ Taking the contribution from $B \to D \tau N$ decays into account, and assuming the unitarity (\ref{unit}) of the matrix $U$, the ratio of branching fractions $R(D)$ is}
\bea \label{RD}
R(D) & \equiv & \frac{ \Gamma(B \to D \tau + {\rm ``missing"}) }{\Gamma(B \to D \ell \nu)}
\nonumber\\
& = & \frac{ \left\{ {\overline \Gamma}(B \to D \tau \nu) + |U_{\tau N}|^2 \left[{\overline \Gamma}(B \to D \tau N) - {\overline \Gamma}(B \to D \tau \nu) \right] \right\} }{  {\overline \Gamma}(B \to D \ell \nu)  } \qquad (\ell =e, \mu) \ ,
\eea
where ${\overline \Gamma}(B \to D \tau \nu)$ is the expression (\ref{bGBDlN}) for zero neutrino mass $M_{\nu}=0$ and $M_{\ell}=M_{\tau}$.
\textcolor{black}{ If the unitarity  is not assumed, $R(D)$ becomes}
\bea \label{RD2}
R(D)  =
\frac{  {\overline \Gamma}(B \to D \tau \nu)
 + |U_{\tau N}|^2  {\overline \Gamma}(B \to D \tau N)  }
 { {\overline \Gamma}(B \to D \ell \nu) }  \quad (\ell =e, \mu) \ .
\eea


\subsection{$R(D^*)$ and decays of $B \to D^* \ell N$}
\label{sub:BtoDstlN}

The matrix elements for $B \to D^*$ transition are more complicated than those for $B \to D$ transition,
because of the vector character of $D^*$, including four form factors:
\bea
H^{\mu}_{\eta}
& = &
i 2 \eta  \frac{\varepsilon^{\mu \nu \alpha \beta}}{(M_B+ M_{\Dst})} \epsilon^*_{\nu} (p_D)_{\alpha} (p_B)_{\beta} V(q^2) - \left[ (M_B+M_{\Dst}) \epsilon^{* \mu} A_1(q^2)
  - \frac{\epsilon^* \cdot q}{(M_B+M_{\Dst})} (p_B+p_D)^{\mu} A_2(q^2) \right]
\nonumber\\
&& + 2 M_{\Dst} \frac{\epsilon^* \cdot q}{q^2} q^{\mu} \left( A_3(q^2) - A_0(q^2) \right) \ ,
\label{FFBDst}
\eea
where
\bes
\label{Hmu}
\bea
H^{\mu}_{(\eta=-1)}
 &\equiv& \langle D^{*-}(p_D) | {\overline c}(1 - \gamma_5) \gamma^{\mu} b | B^{0}(p_B) \rangle
 = \langle {\overline D^{*0}}(p_D) | {\overline c}(1 - \gamma_5) \gamma^{\mu} b | B^{+}(p_B) \rangle
\label{Hetam}
\\
H^{\mu}_{(\eta=+1)}
 &\equiv& \langle D^{*+}(p_D) | {\overline b}(1 - \gamma_5) \gamma^{\mu} c | {\overline B^{0}}(p_B) \rangle
 = \langle D^{*0}(p_D) | {\overline b}(1 - \gamma_5) \gamma^{\mu} c | B^{-}(p_B) \rangle \ ,
\label{Hetap}
\eea
\ees
The four form factors are
\bes
\label{A1VA2}
\bea
A_1(q^2) &=& \frac{1}{2} R_* (w+1) F_*(1) \left[ 1 - 8 \rho_*^2 z(w)
 + (53 \rho_*^2 - 15) z(w)^2 - (231 \rho_*^2 - 91) z(w)^3 \right] \ ,
\label{A1}
\\
V(q^2) &=& A_1(q^2)  \frac{2}{R_*^2 (w+1)} \left[ R_1(1) - 0.12 (w-1) + 0.05 (w-1)^2 \right] \ ,
\label{V}
\\
A_2(q^2) &=& A_1(q^2) \frac{2}{R_*^2 (w+1)} \left[ R_2(1) + 0.11 (w-1) - 0.06 (w-1)^2 \right] \ ,
\label{A2}
\\
A_3(q^2) &=& \frac{(M_B+M_{\Dst})}{2 M_{\Dst}} A_1(q^2) -\frac{(M_B-M_{\Dst})}{2 M_{\Dst}} A_2(q^2) \ .
\label{A3}
\eea
\ees
Here, $R_* = 2 \sqrt{ M_B M_{\Dst}}/(M_B+M_{\Dst})$, the variables $w$ and $z(w)$ are given by Eqs.~(\ref{wz}), and
the values of the free parameters determined in Ref.~\cite{Belle2} are
\bes
\label{paramsDst}
\bea
\rho_*^2 & = & 1.214(\pm 0.035) \ , \qquad 10^3 F_*(1) |V_{cb}| = 34.6(\pm 1.0) \ ,
\label{rhostFst}
\\
R_1(1) & = &1.401(\pm 0.038) \ , \qquad R_2(1) = 0.864(\pm 0.025) \ .
\label{R1R2}
\eea
\ees
We will use the central values of these parameters.

For the decays $B \to D^* \ell N$, the width is given in Ref.~\cite{Cvetic:2016fbv} as
\be \label{GBDstlN}
\Gamma(B \to D^* \ell N) = |U_{\ell N}|^2 \; {\overline \Gamma}(B \to D^* \ell N) \ ,
\ee
where the canonical decay width (i.e., without the heavy-light neutrino mixing) is
\bea
\lefteqn{
{\overline \Gamma}(B \to D^* \ell N) = \frac{1}{64 \pi^3} \frac{G_F^2 |V_{cb}|^2}{M_B^2}
 \int_{(M_N+M_\ell)^2}^{(M_B-M_{\Dst})^2}  d q^2 \;
  \blam^{1/2} |{\vec q}| q^2 {\Bigg \{}
 \left( 1 - \frac{(M_N^2+M_\ell^2)}{q^2} - \frac{1}{3} \blam \right)
 {\bigg [} 2 (M_B+M_{\rm D})^2 A_1(q^2)^2
 }
 \nonumber\\
 &&
 + \frac{8 M_B^2 |{\vec q}|^2}{(M_B+M_{\Dst})^2} V(q^2)^2 +  \frac{M_B^4}{4 M_{\Dst}^2 q^2} \left( (M_B+M_{\Dst})
   \left( 1 - \frac{(q^2+M_{\Dst}^2)}{M_B^2} \right) A_1(q^2) -  \frac{4 |{\vec q}|^2}{(M_B+M_{\Dst})} A_2(q^2) \right)^2 {\bigg ]}
 \nonumber\\
 &&
 + \left[ - \left(\frac{M_N^2-M_\ell^2}{q^2} \right)^2 + \frac{(M_N^2+M_\ell^2)}{q^2} \right] \frac{M_B^2 |{\vec q}|^2}{M_{\Dst}^2 q^2}
 \left[ \frac{2 M_{\Dst} (M_B+M_{\Dst})^2}{\left(2 M_{\Dst} (M_B+M_{\Dst}) - q^2 \right)} \right]^2 \left[ 1 - \frac{(M_B - M_{\Dst}) A_2(q^2)}{(M_B + M_{\Dst}) A_1(q^2)} \right]^2 A_1(q^2)^2 {\Bigg \}},
 \nonumber\\
 \label{bGBDstlN}
 \eea
where $\blam = \lambda \left( 1, \frac{M_N^2}{q^2}, \frac{M_1^2}{q^2} \right)$ and
$|{\vec q}| = \frac{1}{2} M_B \lambda^{1/2} \left( 1, \frac{q^2}{M_B^2}, \frac{ M_{\Dst}^2}{M_B^2} \right)$.
As in the case of $B \to  D l N$, ${\overline \Gamma}(B \to D^* \ell N)$ expresses the SM decay width
$\Gamma_{\rm SM} (B \to D^* \ell \nu_\ell)$, when $N$ is replaced by $\nu_\ell$ (i.e., $M_N \approx 0$).

 \textcolor{black}{ Analogously as in the case of $R(D)$, including the possible contribution from $B \to D^* \tau N$ decays and assuming unitarity (\ref{unit}) of the full $U$ matrix, the ratio of branching fractions $R(D^*)$ can be written as}
\bea \label{RDst}
R(D^*) & \equiv &
\frac{ \Gamma(B \to D^* \tau + {\rm ``missing"}) }{\Gamma(B \to D^* \ell \nu)}
\nonumber\\
& = & \frac{ \left\{ {\overline \Gamma}(B \to D^* \tau \nu)
+ |U_{\tau N}|^2 \left[{\overline \Gamma}(B \to D^* \tau N)
- {\overline \Gamma}(B \to D^* \tau \nu) \right] \right\} }
{ {\overline \Gamma}(B \to D^* \ell \nu) } \qquad (\ell =e, \mu) \ ,
\eea
where ${\overline \Gamma}(B \to D^* \tau \nu)$ is the expression (\ref{bGBDstlN}) for zero neutrino mass $M_{\nu}=0$ and $M_{\ell}=M_{\tau}$.
 \textcolor{black}{ When the matrix $U$ is not assumed to be unitary, $R(D^*)$ becomes}
\bea \label{RDst2}
R(D^*)  = \frac{ {\overline \Gamma}(B \to D^* \tau \nu)
+ |U_{\tau N}|^2 {\overline \Gamma}(B \to D^* \tau N) }
{ {\overline \Gamma}(B \to D^* \ell \nu) } \quad (\ell =e, \mu) \ .
\eea


\subsection{$R(0)$ and decays of  $B \to \ell N$}
\label{BtolN}

%

In the decay $B^+ \to \ell^+ \nu_{\ell}~~(\ell = e, \mu, \tau)$, within the SM with $M_{\nu_{\ell}} \approx 0$,
the decay width is given by
\be
\label{GSMBlnu}
\Gamma_{\rm SM}(B^+ \to \ell^+ \nu_{\ell}) =
 \frac{1}{8 \pi} G_F^2 f_B^2 |V_{ub}|^2 M_B^3 y_{\ell} \left( 1 - y_{\ell} \right)^2  \,
\ee
where $y_{\ell} = M_{\ell}^2/M_B^2$.
Here, $M_B$ and $f_B$ are the $B^+$ meson mass and the decay constant, respectively,
$|V_{ub}|$ is the corresponding CKM matrix element, and
$G_F = 1.1664 \times 10^{-5} \ {\rm GeV}^{-2}$ is the Fermi coupling constant.

The decay width for the process $B^+ \to \ell^+ N$ $(\ell = e, \mu, \tau)$ is, as shown in \cite{Cvetic:2016fbv}:
\be \label{GBlN}
\Gamma(B^+ \to \ell^+ N) =  |U_{\ell N}|^2 ~{\overline \Gamma(B^{\pm} \to \ell^{\pm} N)} \  ,
\ee
where the canonical width ${\overline \Gamma}$ (i.e., without the heavy-light mixing factor $|U_{\ell N}|^2$) is
\be \label{bGBlN}
{\overline \Gamma(B^+ \to \ell^+ N)}
 = {1 \over {8\pi}} G_F^2 f_B^2 |V_{ub}|^2 M_B^3 ~\lambda^{1/2}(1,y_N,y_\ell)~
  \left[ (1 - y_N) y_N + y_\ell (1 + 2 y_N - y_\ell) \right] \ ,
\ee
where $y_N = M_N^2 / M_B^2$ and the function $\lambda^{1/2}$ is given by
\be \label{sqlam}
\lambda^{1/2}(x,y,z) = \left( x^2 + y^2 + z^2 - 2 x y - 2 y z - 2 z x \right)^{1/2} \ .
\ee
It is obvious that if $N$ is replaced by $\nu_\ell$ (i.e., $y_N \approx 0$),
${\overline \Gamma(B^{+} \to \ell^{+} N)}$ in Eq.~(\ref{bGBlN}) becomes
$\Gamma_{\rm SM}(B^+ \to \ell^+ \nu_{\ell})$ in Eq.~(\ref{GSMBlnu}).
And the SM expectation for the newly defined $R(0)$, which is completely independent of hadronic uncertainty,
is given by
\be \label{R0SM}
R(0)_{\rm SM}
 \equiv \frac{ y_{\tau} \left( 1 - y_{\tau} \right)^2 }{ y_{\mu} \left( 1 - y_{\mu} \right)^2 } =(2.2255 \pm 0.0002) \times 10^2 ~,
 \ee
 \textcolor{black}{where the uncertainty comes almost entirely from the uncertainty in the $\tau$ lepton mass \cite{PDG2016}.}

For the decays $B^+ \to \tau^+ +$ ''missing momentum'', taking the contribution from the decays
$B^+ \to \tau^+ N$ into account, the widths can be obtained with the unitarity assumption (5) by
\bea \label{taumissmom}
\Gamma (B^+ \to \tau^+ + {\rm "missing ~momentum"} )
= {\overline \Gamma}(B^+ \to \tau^+ \nu) + |U_{\tau N}|^2 \left[{\overline \Gamma}(B^+ \to \tau^+ N) - {\overline \Gamma}(B^+ \to \tau^+ \nu) \right] \ ,
\eea
where ${\overline \Gamma}(B^+ \to \tau^+ \nu)$ is for zero neutrino mass $M_{\nu}=0$, i.e., it is equal to the SM expression (\ref{GSMBlnu}) with $\ell=\tau$.
%
Considering the contribution from $B^+ \to \tau^+ N$, the ratio $R(0)$ \textcolor{black}{in the considered scenario with one heavy neutrino $N$ and the unitarity (\ref{unit}) of the $U$ matrix is}
\bea \label{R0}
R(0) & \equiv & \frac{ \Gamma(B \to \tau + {\rm ``missing"}) }{\Gamma(B \to \mu \nu)}
\nonumber\\
 &= & \frac{ \left\{ {\overline \Gamma}(B^+ \to \tau^+ \nu)
 + |U_{\tau N}|^2 \left[{\overline \Gamma}(B^+ \to \tau^+ N)
 - {\overline \Gamma}(B^+ \to \tau^+ \nu) \right] \right\} }
 {  {\overline \Gamma}(B^+ \to \mu^+ \nu) } \ .
\eea
Here, ${\overline \Gamma}(B^+ \to \ell^+ N)$ is given in Eq.~(\ref{bGBlN}), and
 ${\overline \Gamma}(B^+ \to \ell^+ \nu)$ is the same expression with zero mass of neutrino $M_{\nu}=0$, i.e., Eq.~(\ref{GSMBlnu}).
On the other hand without the unitarity assumption, we have instead of the relation (\ref{taumissmom}) the following relation:
  \be
\Gamma (B^+ \to \tau^+ + {\rm "missing ~momentum"} )
= {\overline \Gamma}(B^+ \to \tau^+ \nu) + |U_{\tau N}|^2 {\overline \Gamma}(B^+ \to \tau^+ N)  \ ;
\label{taumissmom2}
\ee
and the ratio $R(0)$ becomes
\bea \label{R0_2}
R(0)  = \frac{ {\overline \Gamma}(B^+ \to \tau^+ \nu)
 + |U_{\tau N}|^2 {\overline \Gamma}(B^+ \to \tau^+ N) }
 { {\overline \Gamma}(B^+ \to \mu^+ \nu) } \ .
\eea
As we shall see later, the observation of $R(0)$ can give very useful information on the values of $|U_{\tau N}|$ and the mass of a sterile neutrino $M_N$.


\section{Numerical Analysis and Discussions}
\label{sec:num_disc}

\subsection{$R(D)$, $R(D^*)$ and $B \to D \tau N$, $D^* \tau N$ }
\label{sub:num_RD_RDst}

In this numerical analysis, \textcolor{black}{ first we study $R(D)$ and $R(D^*)$ anomalies in our scenarios: in one scenario  the matrix $U$ without the unitarity assumption; in the other scenario, the full $4 \times 4$  matrix $U$ is considered unitary. As mentioned, the latter scenario is realized when seesaw-type mechanisms are used, and the former may appear when the heavy neutral fermion $N$ originates from a different, unknown, mechanism in a high energy framework beyond the SM.}
Also, we examine the possibility of finding certain ``direct" bounds on magnitudes of the  matrix elements $|U_{\ell N}|$ from this analysis of $R(D)$ as well as $R(D^*)$.

\begin{figure}[h] 
\begin{minipage}[b]{.49\linewidth}
\centering\includegraphics[width=60mm]{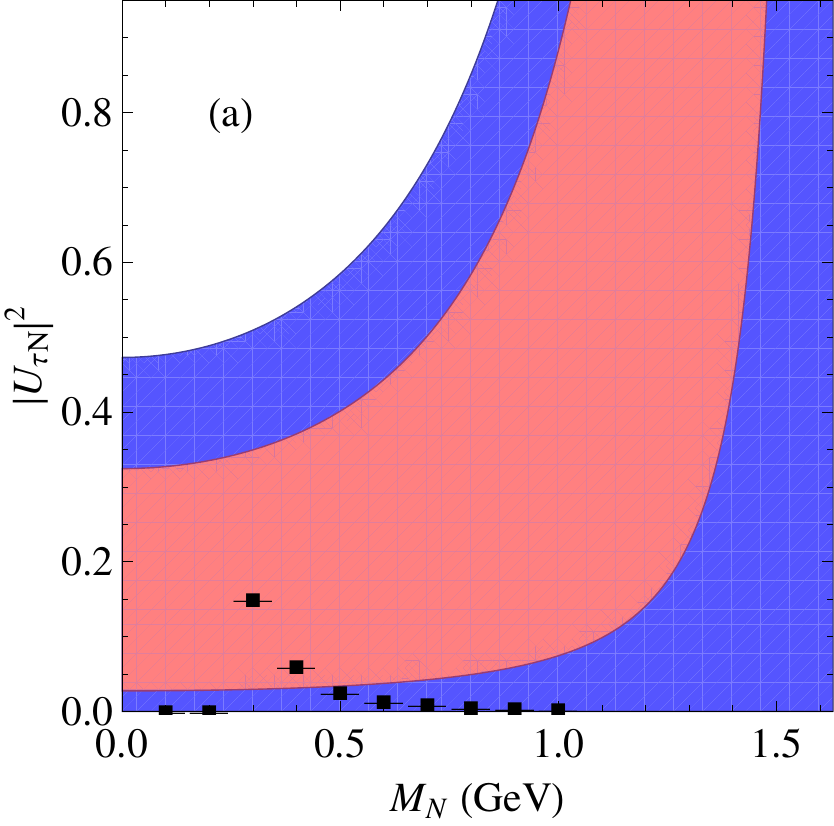}
\end{minipage}
\begin{minipage}[b]{.49\linewidth}
\centering\includegraphics[width=60mm]{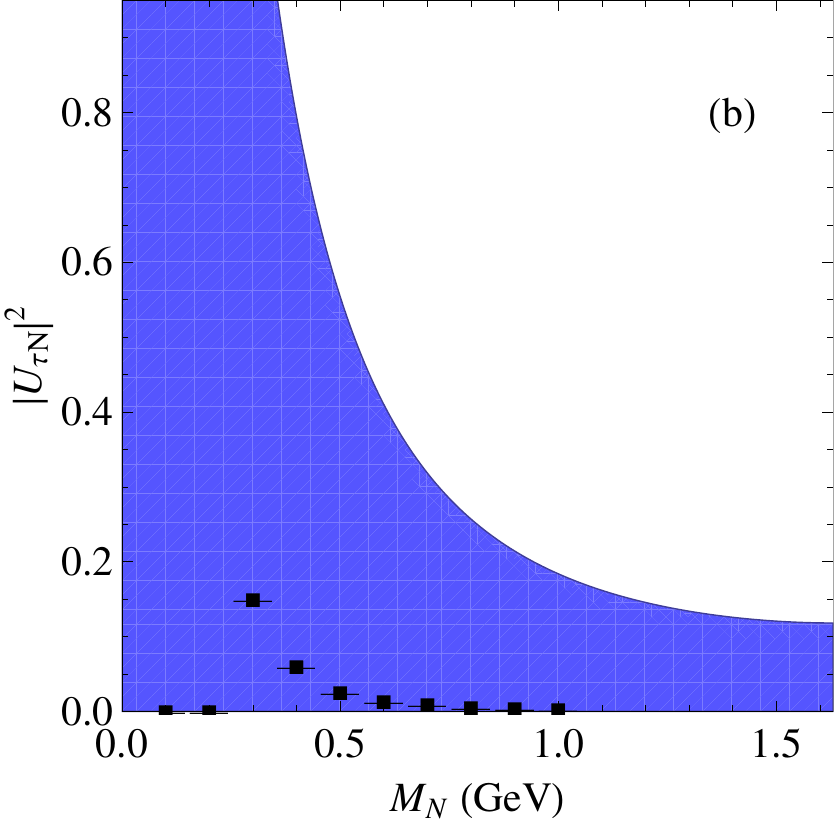}
\end{minipage}
\caption{(color online)  The shaded regions represent the allowed parameter space obtained from the experimental average value of $R(D)$ shown in Table~\ref{tb2}. \textcolor{black}{The red (light grey) region is allowed by experimental data for $R(D)$ at $1\sigma$ level, and the blue (dark grey) region at $2\sigma$ level. The white region is excluded (more than $2\sigma$ deviation): (a) when $U$ is nonunitary, cf.~Eq.~(\ref{RD2}); (b) when $U$ is unitary, cf.~Eqs.~(\ref{unit}) and (\ref{RD}).}
  In (a) and (b), the known
  \textcolor{black}{present upper bounds from CHARM and DELPHI experiments \cite{CHARM,DELPHI}}
  for $|U_{\tau N}|^2$ are denoted by tiny black squares.
\textcolor{black}{In the nonunitary case (a), if accounting for the lepton universality measurements, the three black squares at $M_N^2=0.3$, $0.4$ and $0.5$ GeV decrease to the values $\sim 10^{-2}$ as given in Fig.~\ref{figUtauN2Univ}(a).}
\textcolor{black}{In the unitary case (b), when accounting for the restrictions coming from $\tau \to \ell \nu \nu'$ \cite{Becir}
  \textcolor{black}{and the lepton universality measurements Fig.~\ref{figUtauN2Univ}(b), the black}
  squares at $M_N^2 \geq 0.3$ GeV decrease to the values $|U_{\tau N}|^2 \lesssim 0.01$.}}
\label{figRD}
\end{figure}

\begin{figure}[tb] 
\begin{minipage}[b]{.49\linewidth}
\centering\includegraphics[width=60mm]{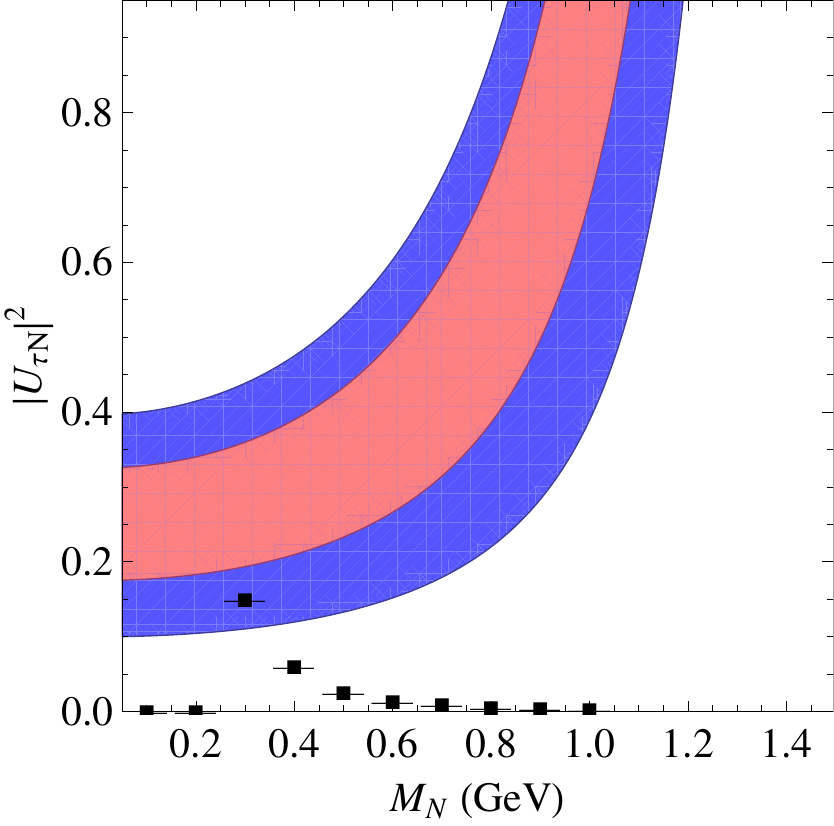}
\end{minipage}
\caption{(color online) The shaded regions represent the allowed parameter space obtained from the experimental average value of $R(D^*)$ shown in Table~\ref{tb2}, in the case of nonunitarity of the full $U$ matrix, cf.~Eq.~(\ref{RDst2}). The red (light grey) region is allowed by the experimental data at $1 \sigma$ level, and the blue (dark grey) region at $2 \sigma$ level. The known \textcolor{black}{CHARM and DELPHI \cite{CHARM,DELPHI} upper bounds} for $|U_{\tau N}|^2$ are denoted by tiny black squares.
\textcolor{black}{If accounting for the lepton universality measurements, the three black squares at $M_N^2=0.3$, $0.4$ and $0.5$ GeV decrease to the values $\sim 10^{-2}$ as given in Fig.~\ref{figUtauN2Univ}(a).}
}
\label{figRDst}
\end{figure}

\begin{enumerate}
\item
We first calculate $R(D)$ including the effect of the process $B \to D \tau N$ as given in Eqs.~(\ref{RD2}) and (\ref{RD}).  By comparing the theoretical result with the experimental data shown in Table~\ref{tb2}, the allowed parameter space for $|U_{\tau N}|^2$ is found in terms of the mass of the sterile neutrino, $M_N$. The results are shown in Fig.~\ref{figRD} for the two scenarios.
\vspace{5pt}

In Fig.~\ref{figRD}(a), the result is found for the scenario without the assumption of unitarity for the $4 \times 4$ matrix $U$.
It shows the allowed parameter space for $|U_{\tau N}|^2$ and $M_N$ obtained from the experimental average value of $R(D)$.  \textcolor{black}{The red (light grey) region is allowed by the experimental data at $1 \sigma$ level, i.e., when the deviation does not surpass $1\sigma$. The blue (dark grey) region is allowed by the data at $2 \sigma$ level, i.e., when the deviation does not surpass $2 \sigma$ (and is above $1 \sigma$). The white region could be regarded as excluded, the deviation there surpasses $2 \sigma$.}
For comparison the known available
\textcolor{black}{upper bounds from CHARM and DELPHI experiments \cite{CHARM,DELPHI}}
for $|U_{\tau N}|^2$ are also included, as tiny black squares\footnote{
  These upper bounds were obtained from various physical processes
\textcolor{black}{searching for heavy sterile neutrino $N$ and are} given in Table~\ref{tb4},
\textcolor{black}{ see footnote \ref{CHDEL} and Refs.~\cite{CHARM,DELPHI,Atre}}.}.
We see that certain range of values of $|U_{\tau N}|^2$ and $M_N$ can fit the experimental data of $R(D)$. At $1 \sigma$ level, there is a tendency that for a smaller value of $M_N$, a smaller $|U_{\tau N}|^2$ can fit the data. For instance, for $M_N = 0.3$ GeV, the smallest value of $|U_{\tau N}|^2$ allowed by the $1 \sigma$ data is $2.8 \times 10^{-2}$, while for $M_N = 1.0$ GeV, the smallest allowed value of $|U_{\tau N}|^2$ is $7.1 \times 10^{-2}$.
\vspace{5pt}

In particular, for $M_N = 0.3$ GeV, the values of $2.8 \times 10^{-2} \lesssim |U_{\tau N}|^2 \lesssim 3.5 \times 10^{-1}$ are allowed by the $1 \sigma$ data, in comparison with the known \cite{CHARM,DELPHI} upper bound $|U_{\tau N}|^2 = 1.5 \times 10^{-1}$.
Similarly, for $M_N = 0.4$ GeV, the $1 \sigma$ data allows the values of $3.0 \times 10^{-2} \lesssim |U_{\tau N}|^2 \lesssim 3.7 \times 10^{-1}$, compared to the DELPHI \cite{DELPHI} upper bound $|U_{\tau N}|^2 = 6.0 \times 10^{-2}$.
We note that for $M_N = 0.3$ GeV and $0.4$ GeV, the smallest allowed value of $|U_{\tau N}|^2$ is smaller than its available \cite{CHARM,DELPHI} upper bound. \textcolor{black}{However, if $M_N \geq 0.5$ GeV,} the values of $|U_{\tau N}|^2$ allowed by the $1 \sigma$ data are larger than the known DELPHI upper bounds.
\textcolor{black}{However, if using the indirect upper bounds coming from the lepton universality measurements, Fig.~\ref{figUtauN2Univ}(a), all the allowed points move out of the $1 \sigma$ region in Fig.~\ref{figRD}(a).}
\vspace{5pt}

In Fig.~\ref{figRD} (b), the result corresponds to the case of considering unitarity of the ($4 \times 4$) matrix $U$, cf.~Eq.~(\ref{unit}).
In this scenario, there is no allowed parameter space for $|U_{\tau N}|^2$ and $M_N$ by the experimental data at $1 \sigma$ level.
At $2 \sigma$ level, certain region of the parameter space is allowed.
For example, for $M_N \lesssim 0.3$ GeV, the values of $0 \lesssim |U_{\tau N}|^2 \lesssim 1$ are allowed, while for $M_N = 0.6$ GeV and $1.0$ GeV, $0 \lesssim |U_{\tau N}|^2 \lesssim 0.4$ and $0 \lesssim |U_{\tau N}|^2 \lesssim 0.2$ are allowed, respectively.
\vspace{5pt}


\item
In the case of $R(D^*)$, it is harder to fit the experimental data by including the contributions from $B \to D^* \tau N$ together with those from the SM processes.
Similarly to the analysis of $R(D)$, we compute $R(D^*)$ including the effect of the decay $B \to D^* \tau N$ as given in Eqs.~(\ref{RDst}) and (\ref{RDst2}), and again examine the two scenarios: \textcolor{black}{(a) one considering the $4 \times 4$ matrix $U$ to be nonunitary, Eq.~(\ref{RDst2}); (b) the other considering $U$ to be unitary, Eqs.~(\ref{unit}) and (\ref{RDst}).}
It turns out that in the latter scenario (b), there are no allowed values of $|U_{\tau N}|^2$ and $M_N$. \textcolor{black}{Namely, Table \ref{tb2} shows that the SM value of $R(D^*)$ is by more than $3 \sigma$ below the central experimental value; on the other hand, the theoretical value in scenario (b) becomes even lower when $U_{\tau N} \not= 0$, cf.~Eq.~(\ref{RDst}).} In contrast, in the former scenario (a), certain values of $|U_{\tau N}|^2$ and $M_N$ \textcolor{black}{ are allowed at $2 \sigma$ level, but not at $1\sigma$ level.} The allowed parameter space is shown in Fig.~\ref{figRDst} for the scenario (a).
For instance, for $M_N = 0.3$ GeV and $0.4$ GeV, the values of $2.0 \times 10^{-1} \lesssim |U_{\tau N}|^2 \lesssim 3.6 \times 10^{-1}$ and $2.1 \times 10^{-1} \lesssim |U_{\tau N}|^2 \lesssim 3.9 \times 10^{-1}$, respectively, \textcolor{black}{ would be allowed at $1\sigma$ level. However, as the tiny black squares in Fig.~\ref{figRDst} indicate, the present \textcolor{black}{CHARM and DELPHI \cite{CHARM,DELPHI} upper bounds} on $|U_{\tau N}|^2$ give compatibility of the experimental $R(D^*)$ with the scenario (a) to at best $2\sigma$ level, and this only if $M_N \approx 0.3$ GeV.}
\textcolor{black}{Further, if we include the indirect upper bounds coming from the lepton universality measurements, Fig.~\ref{figUtauN2Univ}(a), the allowed points move out of the $2 \sigma$ region in Fig.~\ref{figRDst}.}
\end{enumerate}

\vspace{5pt}

\textcolor{black}{We comment on a feature of the results for $R(D^{(*)})$ when the $4 \times 4$ $U$ matrix is unitary.}
To be specific, let us consider the canonical decay widths ${\overline \Gamma}$ for the decays $B \to D^{(*)} \tau N$ given in Eqs.~(\ref{bGBDlN}) and (\ref{bGBDstlN}). \textcolor{black}{We find that $\bar \Gamma (B \to D^{(*)} \tau N) < \bar \Gamma (B \to D^{(*)} \tau \nu)$, i.e.,   $\bar \Gamma (B \to D^{(*)} \tau N)$ is a monotonously decreasing function of $M_N$.}
Thus the $|U_{\tau N}|^2$ term in $R(D^{(*)})$, i.e., in the numerator of Eqs.~(\ref{RD}) and (\ref{RDst}), becomes negative and the effect of this term is to reduce $R(D^{(*)})$ with respect to its SM value. \textcolor{black}{ Only in the scenario where $U$ is nonunitary, cf.~Eqs.~(\ref{RD2}) and (\ref{RDst2}),  does the presence of a heavy neutral fermion $N$ increase the ratio $R(D^{*})$.}


\subsection{$R(0)$ and $B \to \tau N$ }
\label{sub:num_R0}

\begin{figure}[h] 
\begin{minipage}[b]{.49\linewidth}
\centering\includegraphics[width=60mm]{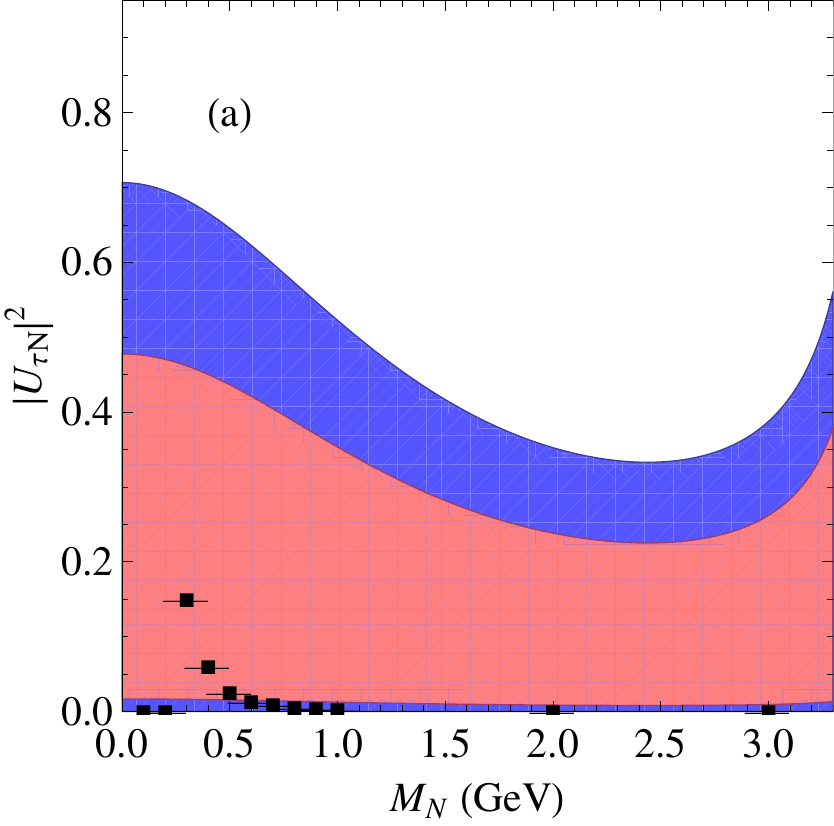}
\end{minipage}
\begin{minipage}[b]{.49\linewidth}
\centering\includegraphics[width=60mm]{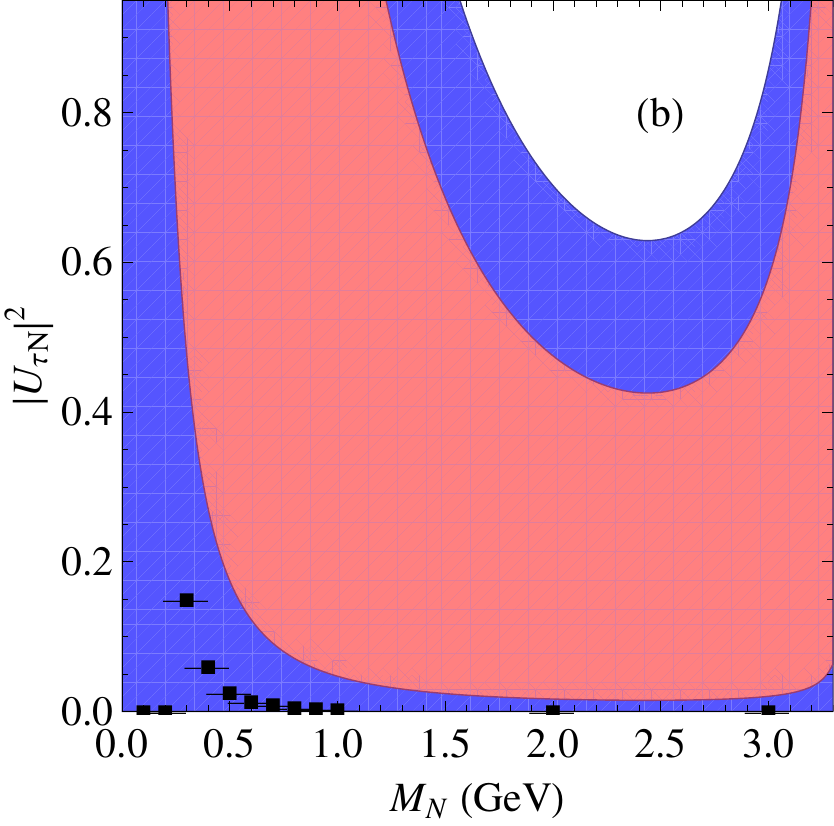}
\end{minipage}
\caption{(color online) The shaded regions represent the allowed parameter space obtained from \textcolor{black}{the experimental average value of ${\rm BF}(B^+ \to \tau^+ \nu) = (1.06 \pm 0.19) \times 10^{-4}$ given in Table~\ref{tb1}. The red (light grey) region is allowed by the experimental data at $1 \sigma$ level, and the blue (dark grey) region at $2 \sigma$ level. Fig.~(a) is obtained by considering $U$ to be nonunitary, Eq.~(\ref{taumissmom2}).
    Fig.~(b) is obtained by considering $U$ to be unitary, Eq.~(\ref{taumissmom}).} In (a) and (b), the known \textcolor{black}{CHARM and DELPHI \cite{CHARM,DELPHI} upper bounds} for $|U_{\tau N}|^2$ are denoted by tiny black squares.
  \textcolor{black}{ If the nonunitary case is generated by a LR-model scenario, we have an additional upper bound $|U_{\tau N}|^2 < 5 \times 10^{-3}$.}
\textcolor{black}{If accounting for the lepton universality measurements, the three black squares at $M_N^2=0.3$, $0.4$ and $0.5$ GeV decrease to the values $\sim 10^{-2}$ as suggested in Figs.~\ref{figUtauN2Univ}(a),(b).}
}
\label{figBtaunu}
\end{figure}

We now analyze the decay process $B \to \tau + {\rm ``missing ~momentum"}$ by including the process $B \to \tau N$.
We will first consider BF of $B \to \tau + {\rm ``missing ~momentum"}$.
\textcolor{black}{ Similarly to the previous cases of $R(D)$ and $R(D^*)$, the two scenarios are examined: (a) one scenario considering the matrix $U$ is nonunitary, and where
  $\Gamma (B^+ \to \tau^+ + {\rm ``missing ~momentum"} )$ is consequently given by Eq.~(\ref{taumissmom2}); (b) the other scenario where the $4 \times 4$ matrix $U$ is considered to be unitary, Eq.~(\ref{unit}), and where $\Gamma (B^+ \to \tau^+ + {\rm ``missing ~momentum"} )$ is consequently given by Eq.~(\ref{taumissmom}).}
The results of this analysis are presented in Figs.~\ref{figBtaunu} (a) and (b).
The allowed values of $|U_{\tau N}|^2$ and $M_N$ are in a wide range for both scenarios.

For future experiments, we make predictions for $R(0)$ given in Eqs.~(\ref{R0}) and (\ref{R0_2}).
They are summarized in Table~\ref{tb4}. The known upper bounds for values of $|U_{\tau N}|^2$,
for various specific values of $M_N$ are taken from Refs. \cite{CHARM,DELPHI,Atre}.
Due to the smallness of these upper bounds for $|U_{\tau N}|^2$, the predicted values of $R(0)$ for various values of $M_N$ do not deviate much from the SM predicted value $R(0)_{\rm SM} = 2.2255 \times 10^{2}$, \textcolor{black}{cf. Eq.~(\ref{R0SM}).}
Only for $0.3 ~{\rm GeV} \leq M_N \leq 0.8 ~{\rm GeV}$, sizable deviations from $R(0)_{\rm SM}$ are expected in the scenario of nonunitarity of $U$. For example, for $M_N = 0.3$ GeV and $0.4$ GeV, the predicted values are $R_0 = 2.5708 \times 10^{2}$ and $2.3672 \times 10^{2}$, respectively.

\begin{figure}[h] 
\begin{minipage}[b]{.49\linewidth}
\centering\includegraphics[width=70mm]{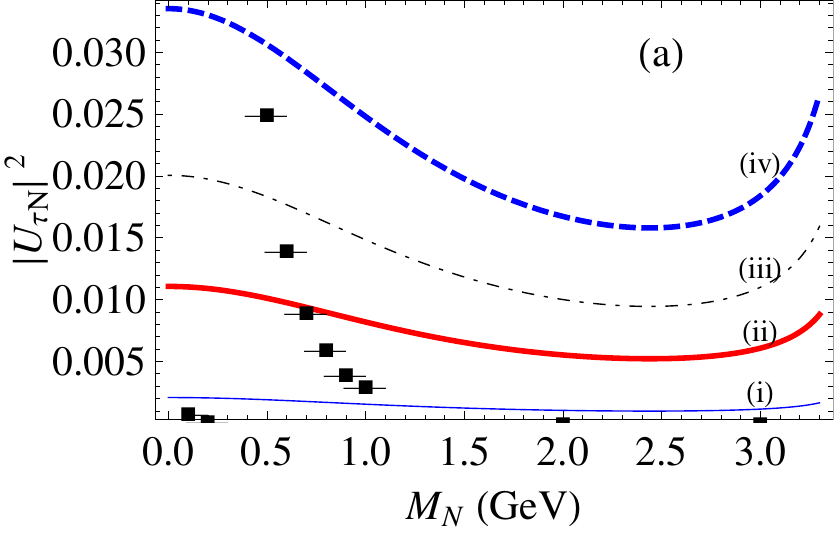}
\end{minipage}
\begin{minipage}[b]{.49\linewidth}
\centering\includegraphics[width=70mm]{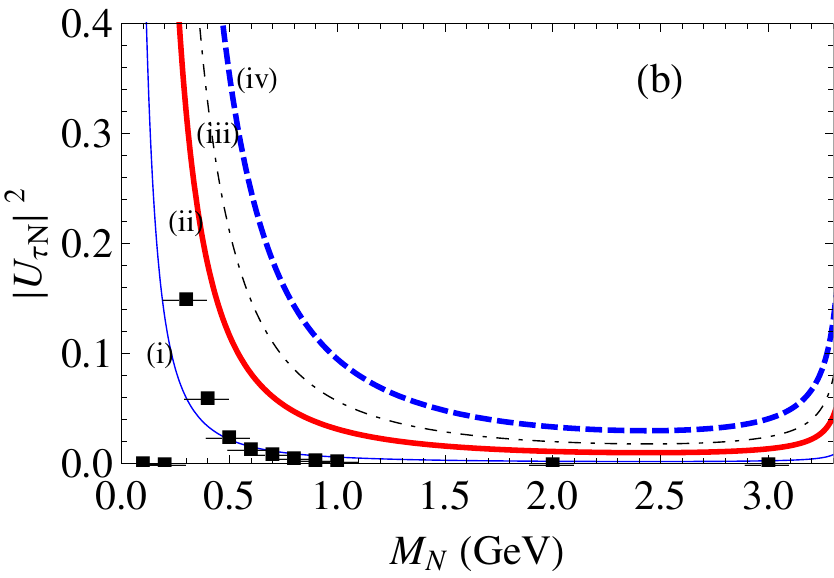}
\end{minipage}
\caption{The graphs of $|U_{\tau N}|^2$ versus $M_N$ when $R_0$ is given.
\textcolor{black}{Fig.~(a) is obtained by considering $U$ to be nonunitary, cf.~Eq.~(\ref{R0_2}).
In Fig.~(b) the $4 \times 4$ matrix $U$ is considered to be unitary, Eqs.~(\ref{unit}) and (\ref{R0}).}
\textcolor{black}{In Figs.~(a) and (b), the cases (i), (ii), (iii), and (iv) correspond to the given value of $R_0 = 2.23 \times 10^{2}$, $2.25 \times 10^{2}$, $2.27 \times 10^{2}$, and $2.30 \times 10^{2}$, respectively.
  In (a) and (b), the known \textcolor{black}{CHARM and DELPHI \cite{CHARM,DELPHI} upper bounds} for $|U_{\tau N}|^2$ are denoted by tiny black squares. }
\textcolor{black}{If the nonunitary case (a) is generated by a LR-model scenario, we have an additional upper bound $|U_{\tau N}|^2 < 5 \times 10^{-3}$.}
\textcolor{black}{If accounting for the lepton universality measurements, according to Figs.~\ref{figUtauN2Univ}, the three black squares at $M_N^2=0.3$, $0.4$ and $0.5$ GeV in Fig.~(a) decrease to the values $\approx 10^{-2}$, and in Fig.~(b) at $M_N \geq 0.3$ GeV the black squares decrease to values $\lesssim 10^{-2}$.}
}
\label{figR0}
\end{figure}

More interesting predictions relevant to $R(0)$ are depicted in Fig.~\ref{figR0} (a) and (b), corresponding to the above two scenarios, respectively. The figures show the graphs of $|U_{\tau N}|^2$ versus $M_N$ for given values of $R(0)$. Provided that the value of $R(0)$ is determined in future experiments (e.g., by measuring the BFs of $B \to \mu \nu$ and $B \to \tau + {\rm ``missing"}$ precisely), one can obtain useful information on $|U_{\tau N}|^2$ and $M_N$ from the figures. For example, if $R(0)$ is determined to be $2.250 \times 10^{2}$, it corresponds to the case (ii) (i.e., red thick line) in Fig.~\ref{figR0} (a) and (b) from which the value of $|U_{\tau N}|^2$ can be extracted for a given $M_N$.
\textcolor{black}{ However, to discriminate experimentally between the cases (i) (SM) and (ii), the branching ratios for $B \to \tau$ + ``missing'' and $B \to \mu \nu$ will have to be measured with precision of $1 \%$ or better; it is possible that such a precision cannot be achieved at Belle-II. Further, if $U$ is nonunitary and if, simultaneously, the $\tau$-$N$-$W$ coupling comes from LR-model scenarios, then we have the upper bound $|U_{\tau N}|^2 < 5 \times 10^{-3}$, cf.~discussion after Eq.~(\ref{U2}). In such a case, we cannot expect to get values of $R(0)$ over $2.24 \times 10^2$ in such scenarios.}
\textcolor{black}{Furthermore, if we take into account the indirect upper bounds on $|U_{\tau N}|^2$ coming from the lepton universality measurements, cf.~Fig.~\ref{figUtauN2Univ}(a), the three black squares at $M_N^2=0.3$, $0.4$ and $0.5$ GeV in Fig.~\ref{figR0} (a) decrease to the values $\approx 10^{-2}$, and  we cannot get values of $R(0)$ over $2.26 \times 10^2$.}

\begin{table}[h]
\small
\centering
\caption{Predicted values of $R_0$:
The $R_0$ values are calculated in two ways: \textcolor{black}{(i) considering the  $U$ to be nonunitary; or (ii) considering $U$ to be unitary. The results are given to four digits to facilitate comparison.}
The values of $|U_{\tau N}|^2$ for various values of $M_N$ are taken
\textcolor{black}{equal to the known CHARM and DELPHI upper bounds  \cite{CHARM,DELPHI}.}
}
\label{tb4}
\begin{tabular}{ c | c | c | c }
\hline \hline
$M_N ({\rm GeV})$ & $|U_{\tau N}|^2$ & ~ $R_0$ ~[{\rm nonunitarity}] ~ & ~ $R_0$ ~[{\rm unitarity}] ~ \\
\hline
0 & 0 & $2.2255 \times 10^{2}$ [SM] & $2.2255 \times 10^{2}$ [SM]   \\
0.1 & $8.0 \times 10^{-4}$ & $2.2273 \times 10^{2}$ & $2.2255 \times 10^{2}$ \\
0.2 & $2.0 \times 10^{-4}$ & $2.2259 \times 10^{2}$ & $2.2255 \times 10^{2}$ \\
0.3 & $1.5 \times 10^{-1}$ & $2.5708 \times 10^{2}$ & $2.2370 \times 10^{2}$ \\
0.4 & $6.0 \times 10^{-2}$ & $2.3672 \times 10^{2}$ & $2.2336 \times 10^{2}$ \\
0.5 & $2.5 \times 10^{-2}$ & $2.2864 \times 10^{2}$ & $2.2307 \times 10^{2}$ \\
0.6 & $1.4 \times 10^{-2}$ & $2.2608 \times 10^{2}$ & $2.2297 \times 10^{2}$ \\
0.7 & $9.0 \times 10^{-3}$ & $2.2491 \times 10^{2}$ & $2.2291 \times 10^{2}$ \\
0.8 & $6.0 \times 10^{-3}$ & $2.2420 \times 10^{2}$ & $2.2286 \times 10^{2}$ \\
0.9 & $4.0 \times 10^{-3}$ & $2.2370 \times 10^{2}$ & $2.2281 \times 10^{2}$ \\
1.0 & $3.0 \times 10^{-3}$ & $2.2345 \times 10^{2}$ & $2.2278 \times 10^{2}$ \\
2.0 & $3.0 \times 10^{-4}$ & $2.2268 \times 10^{2}$ & $2.2262 \times 10^{2}$ \\
3.0 & $4.5 \times 10^{-5}$ & $2.2257 \times 10^{2}$ & $2.2256 \times 10^{2}$ \\
\hline \hline
\end{tabular}
\end{table}

\section{Conclusions}
\label{concl}

In this work we studied the experimental anomalies of the ratios $R(D)$ and $R(D^*)$ related to the semileptonic $B$ decays $B \to D \tau \nu$ and $B \to D^* \tau \nu$, and the newly suggested observable $R(0)$ related to the purely leptonic $B$ decays $B \to \tau \nu$ and $B \to \mu \nu$, considering possible effects from presence of a neutral fermion (sterile neutrino) $N$ with mass $\sim 1$ GeV.
In theoretical estimation of $R(D)$, $R(D^*)$, and $R(0)$, the possible effects of the processes $B \to D \tau N$, $B \to D^* \tau N$, and $B \to \tau N$ were included. For generality we considered two possible scenarios: \textcolor{black}{with the assumption of unitarity of the full $4 \times 4$ mixing matrix $U$ (extended PMNS matrix), and without the assumption of  unitarity.}

We analyzed each of $R(D)$, $R(D^*)$ and $R(0)$ separately. Our findings are summarized as follows.

\begin{enumerate}

\item For the observable $R(D)$, \textcolor{black}{assuming $U$ to be nonunitary,} the discrepancy between the experimental data and the theoretical prediction can be resolved at $1 \sigma$ level [cf.~Fig.~\ref{figRD}(a)] \textcolor{black}{for $M_N < 0.5$ GeV.} The possible values of $|U_{\tau N}|^2$ are found for various values of the mass $M_N$. Especially, we have found that the values of $|U_{\tau N}|^2$ allowed by the $1 \sigma$ data are $2.8 \times 10^{-2} \lesssim |U_{\tau N}|^2 \lesssim 3.5 \times 10^{-1}$ and $3.0 \times 10^{-2} \lesssim |U_{\tau N}|^2 \lesssim 3.7 \times 10^{-1}$ for $M_N = 0.3$ GeV and $0.4$ GeV, \textcolor{black}{which fall within these intervals, respectively.}
These values can be compared with the known
\textcolor{black}{CHARM and DELPHI \cite{CHARM,DELPHI} upper bounds}
$|U_{\tau N}|^2 = 1.5 \times 10^{-1}$ and $6.0 \times 10^{-2}$ for $M_N = 0.3$ GeV and $0.4$ GeV, respectively.
\textcolor{black}{ However, if the nonunitary $U$ has its origin in general $SU(2)_L \times SU(2)_R \times U(1)_{B-L}$ models \cite{LangSan}, these models would imply that $|U_{\tau N}|^2 < 5 \times 10^{-3}$.}
\textcolor{black}{Further, if accounting for the indirect upper bounds on $|U_{\tau N}|^2$ coming from the lepton universality measurements, Fig.~\ref{figUtauN2Univ}(a), the otherwise generous CHARM and DELPHI upper bounds at $M_N \leq 0.5$ GeV get decreased to $|U_{\tau N}|^2 \lesssim 10^{-2}$.}

\item \textcolor{black}{When the full $4\times 4$ matrix $U$ is assumed to be unitary,} in contrast to the above case there is no resolution for the anomaly of $R(D)$ within $1 \sigma$ level, as shown in~Fig.~\ref{figRD}(b).

\item We found it to be more difficult to resolve the anomaly of $R(D^*)$, compared with the $R(D)$ one. The discrepancy in $R(D^*)$ can be resolved only when we assume that $U$ is nonunitary, and only \textcolor{black}{ at best at $2 \sigma$ level} and at $M_N \approx 0.3$ GeV, cf.~Fig.~\ref{figRDst}.
\textcolor{black}{When taking into account the indirect upper bounds on $|U_{\tau N}|^2$ coming from the lepton universality measurements, then the discrepancy cannot be resolved even at $2 \sigma$ level.}

\item We demonstrated that certain useful information on the parameters $|U_{\tau N}|$ and $M_N$ can be extracted from the purely leptonic $B$ meson decays $B \to \tau \nu$, $B \to \tau N$ and $B \to \mu \nu$. If the observable $R(0)$, \textcolor{black}{involving the rates of these decays,} is measured in the future experiments, such as at Belle-II, the value of $|U_{\tau N}|$ could be determined without any hadronic uncertainties, depending on $M_N$, cf.~Fig.~\ref{figR0}.
\end{enumerate}

We assumed that the sterile heavy particle $N$ is stable and invisible, hence does not decay
inside the detector, and thus manifests itself as ``missing momentum'' in the measurements.
However, depending on the values of $U_{lN}$, $M_N$ and the
  detector size, the produced sterile neutrino $N$ can decay within or
  beyond the actual detector.  When $N$
is produced as $B^+ \to \tau^+ N$ or $B^+ \to D^{(*)}\tau^+ N$, and if $N$ also
  decays within the detector, the main signature of $N$ will be
  $N \to l^+ \pi^-$ (if $N$ is Majorana) or
  $N \to l^- \pi^+$  (if $N$ is Dirac or Majorana),
  which will
  appear experimentally as a resonance in $M(l^\pm \pi)$. \textcolor{black}{ Then the experimental signatures would be
  $B \to D^{(*)} l^\pm l^\pm \pi$ and $B \to l^\pm l^\pm \pi$ (if $N$ is Majorana) or
  $B \to D^{(*)} l^\pm l^\mp \pi$ and $B \to l^\pm l^\mp \pi$ (if $N$ is Dirac or Majorana). The details of such decays at Belle-II and LHCb have been
discussed in Ref.~\cite{Cvetic:2016fbv},
for sufficiently small values of $|U_{\ell N}|$, and  $M_N \lesssim 2$ GeV.}


\begin{acknowledgments}
  G.C.~acknowledges the support by FONDECYT (Chile) Grant No.~1130599.
  The work of C.S.K. was supported in part by the NRF grant funded by the Korean government of
  the MEST (No. 2016R1D1A1A02936965).
\end{acknowledgments}

\end{document}